\documentclass[12pt]{article}
\pdfoutput=1
\usepackage{epsfig}
\usepackage{amsfonts}
\usepackage{amssymb}
\usepackage{xcolor}
\usepackage{amsmath}
\usepackage{upgreek}
\usepackage{graphicx}
\usepackage{cite}
\usepackage{calligra}
\usepackage{caption}
\usepackage{subcaption}

\newcommand{\nc}{\newcommand}
\nc{\beq}{\begin{equation}} \nc{\eeq}{\end{equation}}
\nc{\beqa}{\begin{eqnarray}} \nc{\eeqa}{\end{eqnarray}}
\nc{\ba}{\begin{array}} \nc{\ea}{\end{array}}
\begin{document}
\begin{center}

{\bf \LARGE Leading all-loop quantum contribution\\[0.5cm] to the effective potential in general scalar\\[0.5cm] field theory} \vspace{1.0cm}

{\bf \large D. I. Kazakov$^{1,2}$, R.M. Iakhibbaev$^{1}$ \\[0.3cm] and D. M. Tolkachev$^{1,3}$} \vspace{0.5cm}

{\it $^1$Bogoliubov Laboratory of Theoretical Physics, Joint
Institute for Nuclear Research, Dubna, Russia.\\
$^2$Moscow Institute of Physics and Technology, Dolgoprudny, Russia\\ and \\
$^3$Stepanov Institute of Physics, Minsk, Belarus}
\vspace{0.5cm}

\abstract{The RG equation for the effective potential in the leading log (LL) approximation is constructed  which is valid for an arbitrary scalar field theory in 4 dimensions. The solution to this equation sums up the leading $\log\phi$ contributions to all orders of perturbation theory. In general, this is the second order nonlinear partial differential equation, but in some cases it can be reduced to the ordinary one. For particular examples, this equation is solved numerically and the LL effective potential is constructed.  The solution has a characteristic discontinuity replacing the Landau pole typical for the $\phi^4$ theory.  For a power-like potential no new minima appear due to the Coleman-Weinberg mechanism.}
\end{center}


\section{Introduction}
Since the famous paper of S.Coleman and E.Weinberg in 1973 ~\cite{CW} on quantum corrections to the effective potential, the mechanism leading to the appearance of additional minimum due to radiative corrections is called the Coleman-Weinberg mechanism. In their paper, they first considered the scalar field theory
with the potential
\begin{equation}
V_0(\phi)=g\phi^4/4!
\end{equation}
and calculated the one-loop radiative correction which has the form
\beq
V_1(\phi)=\frac{g^2}{16\pi^2}\phi^4\frac{1}{16} \log{\phi^2/\mu^2}.
\eeq
Apparently, at small $\phi$ the $\log{\phi^2/\mu^2}$ is negative and the effective potential $V_{eff}=V_0+V_1$ develops a minimum, as is shown in Fig.1 (left).
\begin{figure}[ht]
\begin{center}
\includegraphics[scale=0.5]{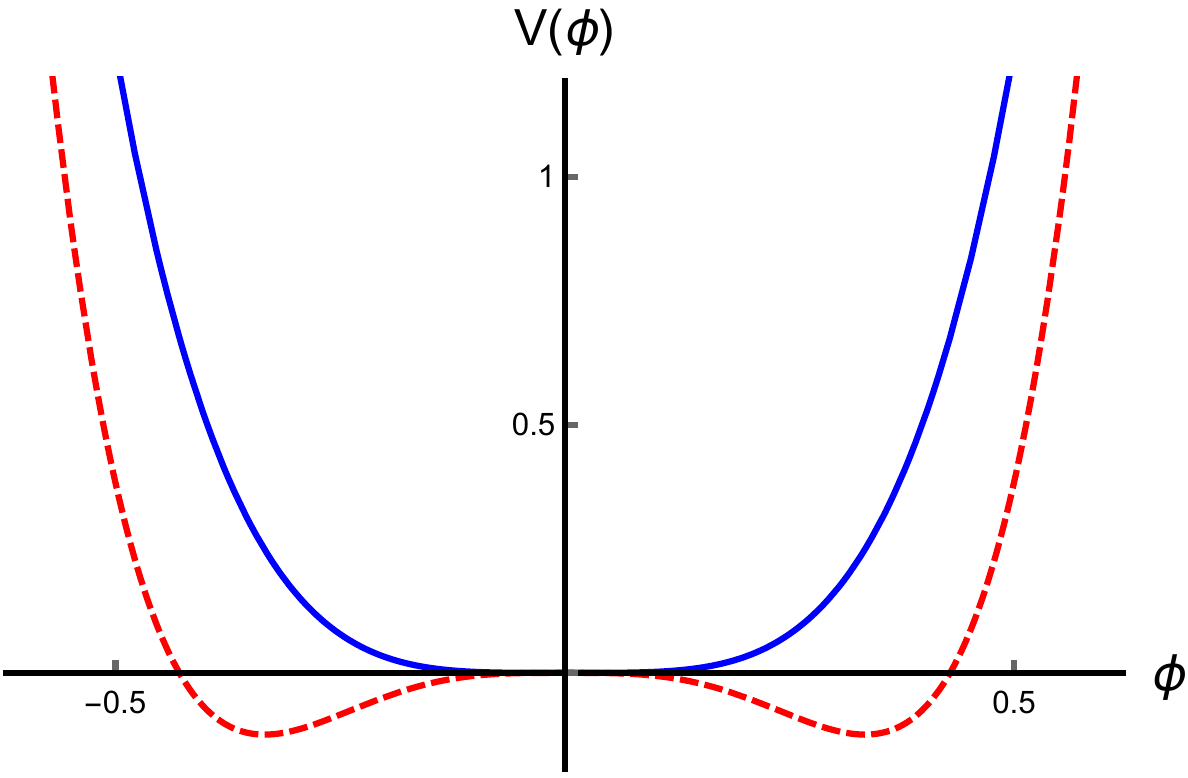} \hspace{1cm}
\includegraphics[scale=0.5]{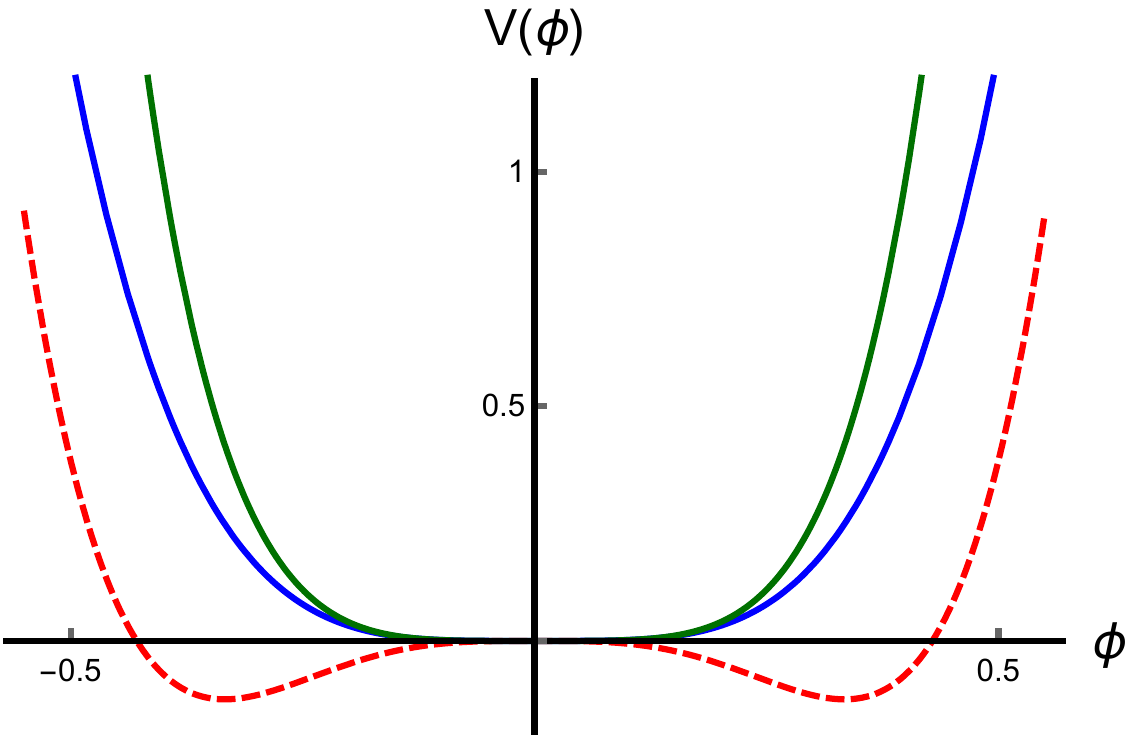}
\end{center}
\caption{Effective potential with the one-loop (left) and leading all-loop (right) radiative corrections}
\label{pot1}
\end{figure}
Thus, a non-trivial minimum appears as a result of the radiative Coleman-Weinberg mechanism. However, as was also mentioned in their paper, if one uses the renormalization group improved effective potential summing up the leading $\log \phi^2/\mu^2$ contributions in all loops, one gets
\beq
V_{eff}^{RG}=\frac{g\phi^4/4!}{1-\frac{g}{16\pi^2}\frac 32\log{\phi^2/\mu^2}}, \label{potential}
\eeq
which obviously has no minimum anymore but has a Landau pole instead, as is shown in Fig.1 (right).

Therefore, in a pure scalar $\phi^4$ theory the proposed C-W mechanism does not work. The authors considered the case of scalar electrodynamics where one has two couplings and they found that for a certain relation between the two couplings the minimum really appears in the renormalization group improved scenario\cite{CW}.

Hence, the role of radiative corrections might be essential, and one has to use the RG method to sum them over all loops. 
Apparently, everything is straightforward in the case of renormalizable interactions and is not obvious otherwise. 
Still, even in the latter case, one can develop a technique
that allows one to sum the leading $\log \phi^2$ corrections in all orders of perturbation theory. Despite the fact that in non-renormalizable interactions one has infinite arbitrariness that cannot be removed by usual renormalization, the leading $\log \phi^2$ terms do not depend on this arbitrariness. This observation justifies the application of the leading log approximation even in the non-renormalizable case. As it happens, these terms can be summed up with the help of generalized RG equations, as will be  shown below.

In this paper, we first derive the generalized RG equation for the LL effective potential which is valid for an arbitrary 4-dimensional scalar field theory and then consider some particular cases where this equation can be reduced to an ordinary differential equation. We solve this equation
numerically and demonstrate the behaviour of the effective potential in the LL approximation.

\section{Effective potential in arbitrary scalar theory in 4 dimensions}

The effective potential is defined as part of the effective action without derivatives. The standard formalism to calculate it is to consider the generating functional for the Green functions in the path integral formalism~\cite{EA}
\beq
Z(J)=\int \mathcal{D} \phi ~ \exp \left(i\int d^4x ~ {\cal L}(\phi,d \phi)+J\phi \right),
\eeq
which corresponds to vacuum-to-vacuum transitions in the presence of the classical external current $J(x)$. It is convenient further to take the  functional $W(J)=-i\log Z(J)$ responsible for generating connected correlation functions.

Now the quantum effective action is obtained using the Legendre transformation of $W(J)$
\beq
\Gamma(\phi)= W(J)-\int d^4x J(x)\phi(x),
 \eeq
where the classical field $\phi(x)$ is defined as the solution to $\phi(x)=\frac{\delta W(J)}{\delta J(x)}$.

A direct way to calculate the effective action $\Gamma(\phi)$ perturbatively  is to sum over all 1PI vacuum diagrams acquired using the Feynman rules derived from the shifted action $S[\phi+\widehat \phi]$, where $\phi$ is the classical field obeying the equation of motion and $\widehat \phi(x)$ is the quantum field~\cite{EA}.
 
\begin{figure}[ht]
 \begin{center}
  \epsfxsize=10cm
 \epsffile{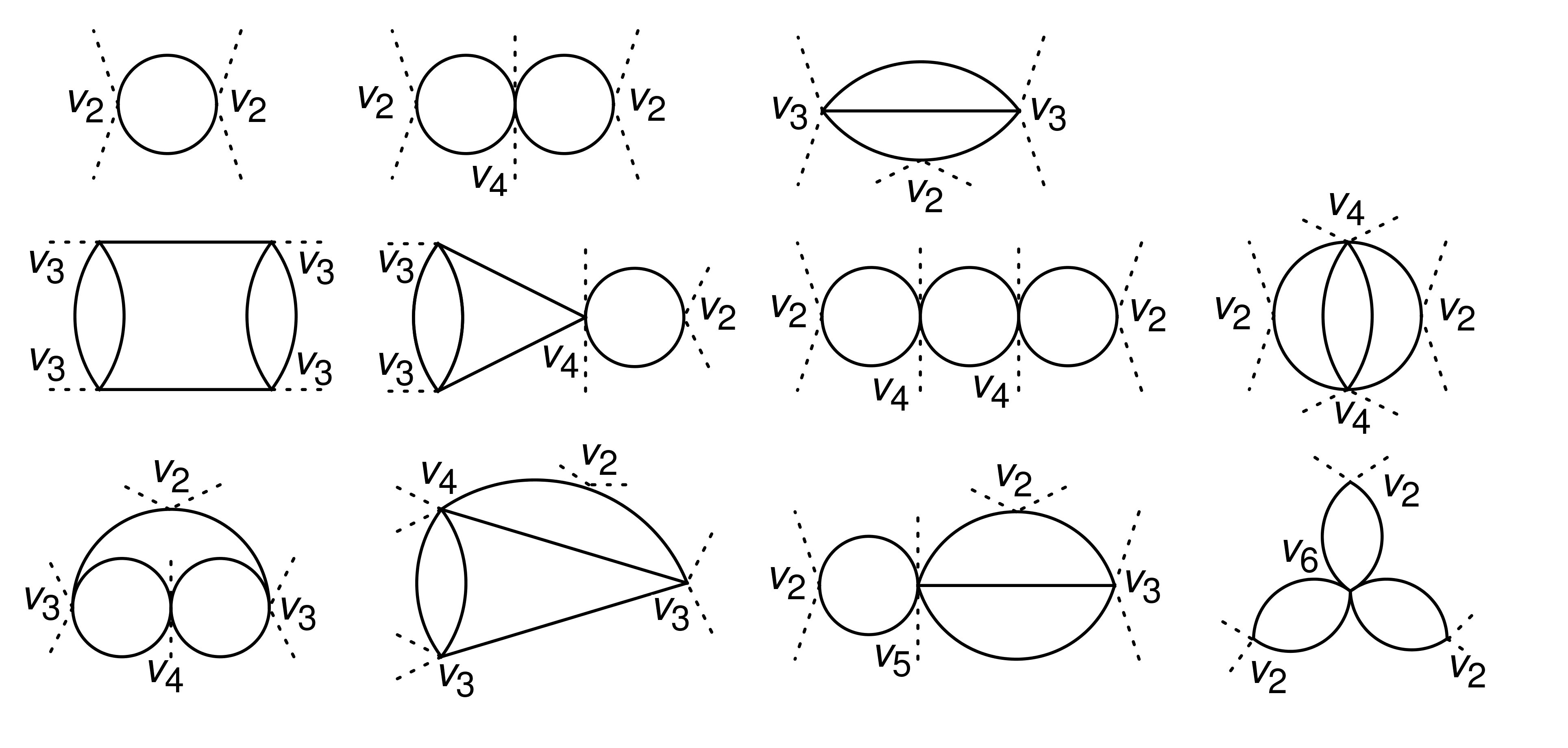}
 \end{center}
 \vspace{-0.2cm}
 \caption{The one-, two- and three-loop vacuum graphs contributing to the effective potential} 
\label{vacgr}
 \end{figure}

Converting this prescription into the form of Feynman diagrams for a scalar field theory with the Lagrangian
\beq
{\cal L} = \frac 12 (\partial_\mu \phi)^2-  gV_0(\phi)
\eeq
one has to take the 1PI vacuum diagrams with the propagator containing  an infinite number of insertions 
of $v_2(\phi)\equiv \frac{d^2V_0(\phi)}{d\phi^2}$ which
acts like a mass term which, however, depends on the field $\phi$: $m^2(\phi)=gv_2(\phi)$. The vertices are also generated by expansion of $V_0(\phi+\widehat \phi)$ over $\widehat\phi$.  After that we calculate the effective potential in a power series over the coupling $g$, so that
\beq
V_{eff}=g\sum_{n=0}^\infty (-g)^n V_n. 
\eeq
The one-, two- and three-loop vacuum diagrams contributing to the effective potential are presented in Fig.\ref{vacgr}, where the classical external fields are symbolically shown by dotted lines. The number of external lines in each vertex is not fixed and depends on the choice of the classical potential, namely it is given by derivatives of the potential $v_n\equiv d^n V_0/d\phi^n$. The number of derivatives  in  its turn corresponds to the number of internal quantum lines in each vertex as is shown explicitly in Fig.\ref{vacgr}.  

The next step is to calculate the vacuum diagrams and to extract the leading $\log\phi$ behaviour. To do this, we notice that all the vacuum diagrams are UV divergent and the $\log\phi$ behaviour
is linked to these divergences. In what follows, we regularize the UV divergences with the help of dimensional regularization taking the dimension equal to $d=4 -2 \epsilon$. Then the task is to evaluate the leading $1/\epsilon^n$ pole terms in each diagram in the n-th order of PT.  The coefficients of the $\log^n\phi$ terms coincide with those of the $1/\epsilon^n$ terms. The  easiest way to do it is to consider the  dimensionless logarithmically divergent diagrams as those shown in Fig.\ref{vacgr}.  

Then, for the one-loop diagram one has the following expression:
\beq
V_1= \frac 14\frac{v_2^2}{\epsilon} \left(\frac{\mu^2}{m^2}\right)^\epsilon \to \frac{v_2^2}{4}\left(\frac 1\epsilon +\log \frac{\mu^2}{m^2}\right), \ \ \  m^2=gv_2(\phi),
\eeq
where 1/4 is the combinatoric factor.
The coefficient of the leading logarithm is always equal to the one of the leading pole.
Thus, replacing the poles by $1/\epsilon \to -\log g v_2/\mu^2$, one gets the  desired  leading $\log\phi$ behaviour. For the potential $V_0(\phi)=\phi^4/4!$ the mass is $m^2(\phi)=g\phi^2/2$
and $\log \frac{m^2}{\mu^2}=\log \frac{g\phi^2}{2\mu^2}$. It is more complicated otherwise.

As was already mentioned external legs contain derivatives of the original potential $v_n$. So that according to Fig.\ref{vacgr} at one loop one has $V_1\sim v_2^2$, at two loops $V_2\sim v_2^2v_4+v_3^2v_2$, etc. 
For the $\phi^4$ theory all these terms eventually are equal to $\phi^4$ and divergences are cancelled by the standard counter term $(Z_4-1)\phi^4$. However, for an arbitrary potential this is not the case. For instance, for the $\phi^6$ theory $v_2\sim \phi^4, v_3\sim \phi^3, v_4\sim \phi^2$, etc and one has increasing powers of the fields at each new order of PT. These divergences can be absorbed into  counter terms with higher powers of the fields which are not present in the original Lagrangian. This is a characteristic feature of non-renormalizable interactions. Introducing  these counter terms
one has increasing arbitrariness that in general cannot be removed by the usual renormalization of masses and couplings. In what follows we do not pretend to solve this major problem but assume that divergences are subtracted somehow, for instance, by the minimal subtraction.  Below we consider the action of the ${\cal R}$-operation that subtracts the UV divergences. Fortunately, the coefficients of the leading poles (and leading logarithms) do not depend on this arbitrariness, they are universal! This observation justifies the application of the leading log approximation even in the non-renormalizable case.  Thus, in non-renormalizable theories, even when one cannot control the divergences, there are some quantities like the leading logs of momenta for the scattering amplitudes or the leading logs of the field in the case of effective potential, that can be evaluated and obey the corresponding RG equations. The derivation of these equations is the aim of the present paper.

The evaluation of the poles from the n-loop Feynman diagrams is not an easy task in itself but if one is interested in leading poles, it can be reduced to the one-loop divergences, as it follows from the local nature of the BPHZ renormalization operation~\cite{BP,Hepp,Zimmermann}. This is equally true for renormalizable as well as non-renormalizable interactions and allows us to calculate the leading poles in a pure algebraic way without really evaluating the complicated Feynman integrals. We described this procedure in detail in a series of our publications~\cite{we2015, we2016, we2017, Kazakov:2019wce ,we2022} and will reproduce here only the summary.

\section{RG equation for effective potential in arbitrary scalar theory in 4 dimensions}

\begin{figure}[ht]
 \begin{center}
  \epsfxsize=13cm
 \epsffile{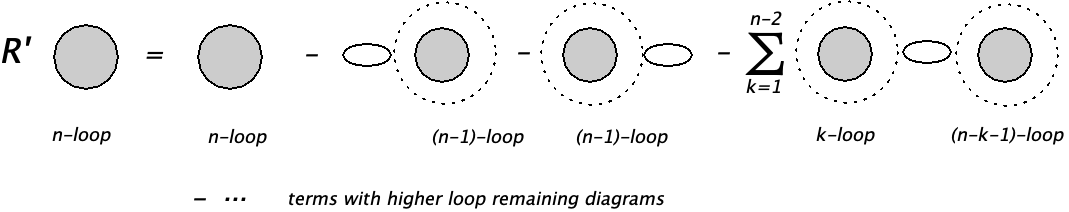}
 \end{center}
 \vspace{-0.2cm}
 \caption{$\mathcal{R}'$-operation for the leading divergences} 
\label{RopEff}
 \end{figure}

Remind that the ${\cal R}$-operation~\cite{BogoliubovBook}  acting on a n-loop diagram subtracts first of all the UV divergences in subgraphs starting from one loop and up to (n-1) loops and then finally subtracts the remaining n-loop divergence which is always local due to the Bogoliubov-Parasyuk theorem~\cite{BP,Hepp,Zimmermann}. This n-loop divergence left after the incomplete ${\cal R'}$-operation is precisely what we are looking for. Then the locality requirement tells us that the leading $1/\epsilon^n$ divergence at n-loops $A^{(n)}_n$ is given by the following
formula~\cite{we2015}:
\beq
A^{(n)}_n=(-1)^{n+1}\frac 1n A^{(1)}_n, \label{red}
\eeq
where $A^{(1)}_n$ is the one-loop divergence left after subtraction of the $(n-1)$-loop counter term as a result of the incomplete ${\cal R'}$-operation. Recall that the $\mathcal{R'}$-operation for any graph $G$ can be defined  recursively via the action of the $\mathcal{R'}$-operation on divergent subgraphs~\cite{Vasiliev, Collins}:
\begin{equation}
    \mathcal{R}'G= \left(1- \sum_\gamma K\mathcal{R}'_\gamma+\sum_{\gamma \gamma' } K\mathcal{R}'_\gamma K\mathcal{R}'_{\gamma'} - \ldots  \right) G,
\end{equation}
where the subtraction operator $K_\gamma$ subtracts the UV divergence of a given subgraph $\gamma$. Graphically, it is  presented in Fig.\ref{RopEff}, 
where the subgraphs inside the dotted line correspond to the action of the $K_\gamma$-operator and have to be shrunken to a point. The live higher loop  subgraphs  obtained throughout the  application of the $\mathcal{R'}$-operation are irrelevant for our purposes due to eq.(\ref{red}).

Thus, the main 'building block' of the entire incomplete $\mathcal{R}'$-operation is the one loop live diagram that stands either to the left or to the right or the middle of the diagram, as shown in Fig.\ref{RopEff}.

\begin{figure}[ht]
 \begin{center}
  \epsfxsize=12cm
\epsffile{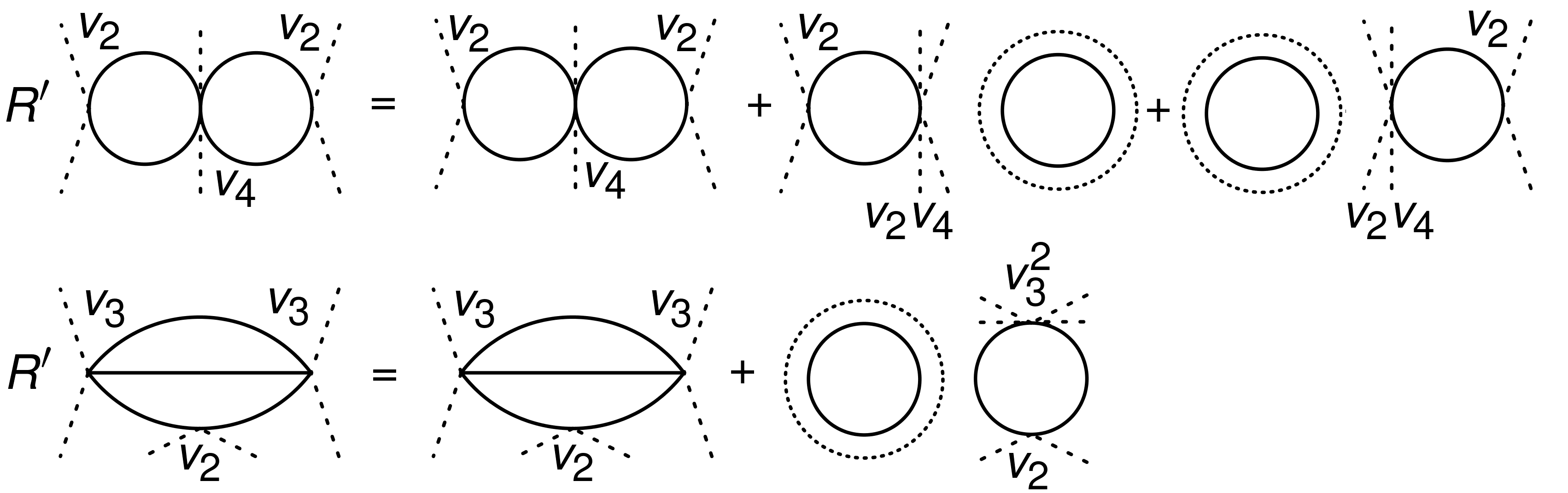}
 \end{center}
 \vspace{-0.2cm}
 \caption{$\mathcal{R}'$-operation for the two-loop diagrams} 
\label{R-op}
 \end{figure}

For the illustration of how the $\mathcal{R}'$-operation works we consider the two loop diagrams from Fig.\ref{vacgr}. The action of the incomplete $\mathcal{R}'$-operation results in subtraction of the one loop subgraphs as shown in Fig.\ref{R-op}. 

The loop inside the dotted line denotes the singular part of the one loop divergent subgraph. This subtraction corresponds to taking into account the one loop counter term $\Delta V_1=1/4 v_2^2/\epsilon$.  To get the new vertices that appear in the subtracted diagrams in Fig.\ref{R-op} one has to make the same shift of the argument in the one loop counter term $\Delta V_1$ as in the original Lagrangian, namely to take $\Delta V_1(\phi+\widehat \phi)$ and expand it up to the second power of $\widehat \phi$. One gets 
\beq \Delta V_1(\phi+\widehat \phi)=\frac14 v_2^2(\phi+\widehat \phi)/\epsilon=\frac 14(v_2^2+2v_2v_3\widehat \phi+v_3^2\widehat \phi^2+2v_2v_4\widehat \phi^2+...)/\epsilon, \label{exp}
\eeq
that gives us exactly the needed new vertices $\frac 14 v_3^2\widehat\phi^2/\epsilon$ and $\frac 12v_2v_4\widehat\phi^2/\epsilon$. The linear term does not work since it corresponds to the vacuum term equal to zero.  
  
 To calculate the needed $1/\epsilon^2$ term for the two loop diagrams one can now use the $\mathcal{R}'$-operation. Namely, for the diagrams shown in Fig.\ref{R-op} one has (hereafter we consider only the leading divergences)
 \beqa
 \mathcal{R}'Diag_1&=& \frac{A_2^{(2)}}{\epsilon^2}\left(\frac{\mu^2}{m^2}\right)^{2\epsilon}v_2^2v_4-2\frac{A_1^{(1)}}{\epsilon}\left(\frac{\mu^2}{m^2}\right)^\epsilon v_2^2v_4\frac{A_1^{(1)}}{\epsilon},\label{d1}\\
  \mathcal{R}'Diag_2&=& \frac{A_2^{(2)}}{\epsilon^2}\left(\frac{\mu^2}{m^2}\right)^{2\epsilon}v_3^2v_2-\frac{A_1^{(1)}}{\epsilon}\left(\frac{\mu^2}{m^2}\right)^\epsilon v_3^2v_2\frac{A_1^{(1)}}{\epsilon}.\label{d2}
 \eeqa
 Cancellation of the singular terms $log(\mu^2/m^2)/\epsilon$, required by locality of the $\mathcal{R}'$-operation, leads to a simple relation $A_2^{(2)}=(A_1^{(1)})^2$ for the first diagram and $A_2^{(2)}=1/2(A_1^{(1)})^2$ for the second one, thus reducing the leading two loop divergence to the one loop case. Since $A_1^{(1)}=1$, one has $A_2^{(2)}=1$ and $A_2^{(2)}=1/2$, for the first and the second diagrams, respectively, thus giving the total  two loop contribution
 \beq
 \Delta V_2=\frac 18(v_2^2 v_4+v_3^2v_2)/\epsilon^2, \label{2loop}
 \eeq 
 where we take into account the combinatorial factors of 1/8 and 1/4, respectively. Notice again, that the coefficients of $\log^2(\mu^2/m^2)$ from the diagrams of eqs.(\ref{d1},\ref{d2}) coincide with those of the 
 $1/\epsilon^2$ pole as one can deduce expanding eqs.(\ref{d1},\ref{d2}) over $\epsilon$,
 namely
 \beqa
 &&\frac{A_2^{(2)}}{2}4\log^2\left(\frac{\mu^2}{m^2}\right)-2\frac{(A_1^{(1)})^2}{2}\log^2 \left(\frac{\mu^2}{m^2}\right)=(A_1^{(1)})^2\log^2\left(\frac{\mu^2}{m^2}\right)=\log^2\left(\frac{\mu^2}{m^2}\right),\\
 &&\frac{A_2^{(2)}}{2}4\log^2\left(\frac{\mu^2}{m^2}\right)-\frac{(A_1^{(1)})^2}{2}\log^2\left(\frac{\mu^2}{m^2}\right)=\frac{(A_1^{(1)})^2}{2}\log^2\left(\frac{\mu^2}{m^2}\right)=\frac 12\log^2\left(\frac{\mu^2}{m^2}\right).
 \eeqa
 
 The same way the $\mathcal{R}'$-operation  works in higher orders, the counter terms of the previous order have to be shifted and expanded to get the corresponding new vertices. On the level of the individual diagrams one have just to subtract the  appeared divergent  subgraphs. The leading divergences then can be calculated using the $\mathcal{R}'$-operation  as shown in Fig.\ref{RopEff}.  As an illustration we show also the three loop result obtained from the diagrams shown in Fig.\ref{vacgr}. One has
 \beq 
 \Delta V_3=\frac{v_3^4}{48 \epsilon ^3}+\frac{3 v_2 v_4 v_3^2}{16 \epsilon ^3}+\frac{v_2^2 v_5 v_3}{8 \epsilon ^3}+\frac{5 v_2^2 v_4^2}{48 \epsilon ^3}+\frac{v_2^3 v_6}{48 \epsilon ^3}. \label{3loop}
 \eeq

The next step is to define the recurrent structure  of the leading divergences. According to Fig.\ref{RopEff}, if the one loop diagram stands to the left or to the right of the whole diagram, it has two vertices that have a pair of quantum lines: one is $v_2$ and the other is obtained from the $(n-1)$-loop  counter term expression $\Delta V_{n-1}(\phi)$ by the same shift of the argument $\phi \to\phi+ \widehat\phi$ and expanding up to the second power of $\widehat\phi$ as in the case of one loop (\ref{exp}). This gives $D_2  \Delta V_{n-1}\equiv d^2 \Delta V_{n-1}/d\phi^2$. In the case when the one-loop diagram stands in the middle, on has two similar vertices $D_2  \Delta V_{k}$. As a result of eq.(\ref{red}) and taking into account the combinatorics, the recurrence relation looks like
\beq
n \Delta V_n= \frac 12 v_2 D_2  \Delta V_{n-1} + \frac 14 \sum_{k=1}^{n-2} D_2  \Delta V_k D_2  \Delta V_{n-1-k}, \ \ \ n\geq 2 \label{rec}
\eeq
with the boundary value $\Delta V_1=1/4v_2^2$.  This term can actually be included in the last term of eq.(\ref{rec}) defining  $\Delta V_0=V_0$, and one gets
\beq
n \Delta V_n=  \frac 14 \sum_{k=0}^{n-1} D_2  \Delta V_k D_2  \Delta V_{n-1-k}, \ \ \, n\geq1, \ \   \Delta V_0=V_0 .  \label{rec2}
\eeq
One can check that the two- and three loop expressions (\ref{2loop},\ref{3loop}) can be reproduced from recurrence relation (\ref{rec2}) by explicit differentiation.

Remind that $\Delta V_n$ is the coefficient of the leading pole in PT expansion; hence, the  expansion parameter is actually $z=\frac{g}{\epsilon}$.  Using the recurrence relation (\ref{rec2}), one can generate the leading divergence in every order of PT pure  algebraically. One can then define the sum
\beq
\Sigma(z,\phi)=\sum_{n=0}^{\infty} (-z)^n\Delta V_n(\phi),
\eeq
that obeys the equation  which can be obtained from (\ref{rec2}) multiplying it by $(-z)^{n-1}$ and summing from 
$n=1$ to infinity. One gets the main equation valid for an arbitrary potential
\beq
\boxed{\frac{d \Sigma}{dz}=-\frac 14 (D_2 \Sigma)^2, \ \ \Sigma(0,\phi)=V_0(\phi)}. \label{RG}
\eeq
This is in fact the desired generalized RG pole equation.  To get the effective potential
one has just make the substitution
\beq
V_{eff}(g,\phi)=g\Sigma(z,\phi)|_{z\to -\frac{g}{16\pi^2}\log{gv_2/\mu^2}}.
\eeq
Remind that  $v_2(\phi)\equiv \frac{d^2V_0(\phi)}{d\phi^2}$.

Notice that eq.(\ref{RG}) is in fact a partial differential equation since the function $\Sigma(z,\phi)$ depends on two variables: $z$ and $\phi$. However, in some cases it can be reduced to an ordinary differential equation which can be solved at least numerically. We consider these examples below.

\section{Example I: Power-like potential}

Consider the power-like potential
\beq
gV_0(\phi)=g\frac{\phi^p}{p!}.
\eeq
In this case, the  mass dimension of the coupling $g$ is $[4-p]$ and a dimensionless combination 
will be $g\phi^{p-4}$. This means that the function $\Sigma(z,\phi)$ can be represented
 in the following form:
\begin{equation}
 \Sigma(z,\phi)= \frac{\phi^p}{p!} f(z \phi^{p-4}). \label{sigma}
\end{equation}
Substituting this form of $\Sigma(z,\phi)$ into eq.(\ref{RG}) and introducing a new dimensionless variable
$y=z \varphi^{p-4} $,  one can reduce the main equation to the ordinary differential one: 
\begin{equation}
    f'(y)=-\frac{1}{4p!} \left[p (p-1) f(y)+(p-4)(3p-5) y f'(y) + (p-4)^2 y^2 f''(y)\right]^2
    \label{phiPeq}\end{equation}
with the initial conditions being 
\begin{equation}
    f(0)=1, f'(0)=-\frac{1}{4} \frac{p(p-1)}{(p-2)!}.
\end{equation}
This equation escapes analytical solution but can be solved numerically. Below we consider some particular cases.

\subsection{p=4}
In this case, eq.(\ref{phiPeq}) is simplified and one gets
\beq
    f'(y)=-\frac{3}{2}f(y)^2.
    \label{phi4eq}
\eeq
One can recognise the usual RG equation for the effective coupling in the $g\phi^4$ theory with the one-loop beta-function coefficient $3/2$. It can be solved analytically with the result:
\begin{equation}
    f(y)=\frac{1}{1+\frac{3}{2} y}.
\end{equation}
Thus, the solution is a simple geometrical progression which has a pole at $y= -3/2$. Replacing now the argument $y \to   -g/16\pi^2 \log(\frac{g\phi^2}{2\mu^2})$, one obtains the  effective potential in the LL approximation
\begin{equation}
    V_{eff}(\phi)=\frac{g \phi^4/4!}{1-\frac{3}{2} \frac{g}{16\pi^2} \log \left(\frac{g \phi^2}{ 2\mu^2}\right)}
\end{equation}
in agreement with eq.({\ref{potential}) above. It has the form  of a potential well with walls of infinite height which corresponds  to the pole at  $g/16\pi^2 \log(\frac{g\phi^2}{2\mu^2}) =2/3$, as can be seen in Fig.\ref{EffPot4}.
\begin{figure}[h!]
 \begin{center}
  \epsfxsize=8cm
 \epsffile{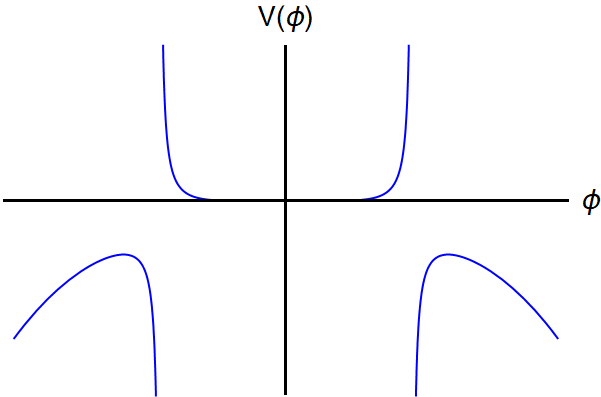}
 \end{center}
 \vspace{-0.2cm}
 \caption{Resummed $\phi^4$-potential} 
\label{EffPot4}
 \end{figure}
 
The validity of perturbation theory in this case requires the expansion parameter $g/16\pi^2 <1$ and the validity of the LL approximation requires  $ \log(\frac{g\phi^2}{2\mu^2})>1$.  These conditions can be easily satisfied for small coupling $g$ and large field $\phi$; however, the region of the pole is already on the edge of validity of PT.

\subsection{p$>$4}

It can be seen that eq.(\ref{phiPeq}) in general has a higher non-linearity for all cases when $p>4$. At the same time, one can notice that the right-hand side is homogeneous, while the left-hand side is not. The situation is similar to the problem  of motion in the Coulomb potential~\cite{Coulomb}, where the equation has an analogous form. Also, there appear singularities which turn out to be avoidable by choosing convenient coordinate transformations or modifications of the solution.  This is also similar to our eq.(\ref{phiPeq}) where  it is useful to choose the function $f(y)$ in the form $f(y)=  u(y)/y$. Such a substitution improves the behaviour of the solution near the singularity, the function $u(y)$ is smooth near $y=0$ and is suitable for numerical calculations. 

Replacing $f$ by $u$, we come to the equation
\begin{equation}
 yu'(y)-u(y)=-\frac{1}{4p!}[12 u(y)+(p-4)(p+3)yu'(y)+(p-4)^2y^2u''(y)]^2,  
       \label{uequation}
\end{equation}
which is now totally homogeneous and possesses the reflection symmetry $x\to -x$.  Its solution  with the initial condition $u(\pm 0)=0, u'(\pm0)=\pm1$ is symmetric with respect to  $x\to -x$ and has no singularity at $y=0$.  Qualitatively, the form of the solution for $u(y)$ and $f(y)$ is presented in Fig.\ref{sol}. One can see that the function $f$ has finite discontinuity at $y=0$ which translates into discontinuity in the effective potential.
\begin{figure}[ht]
\begin{center}
\includegraphics[scale=0.45]{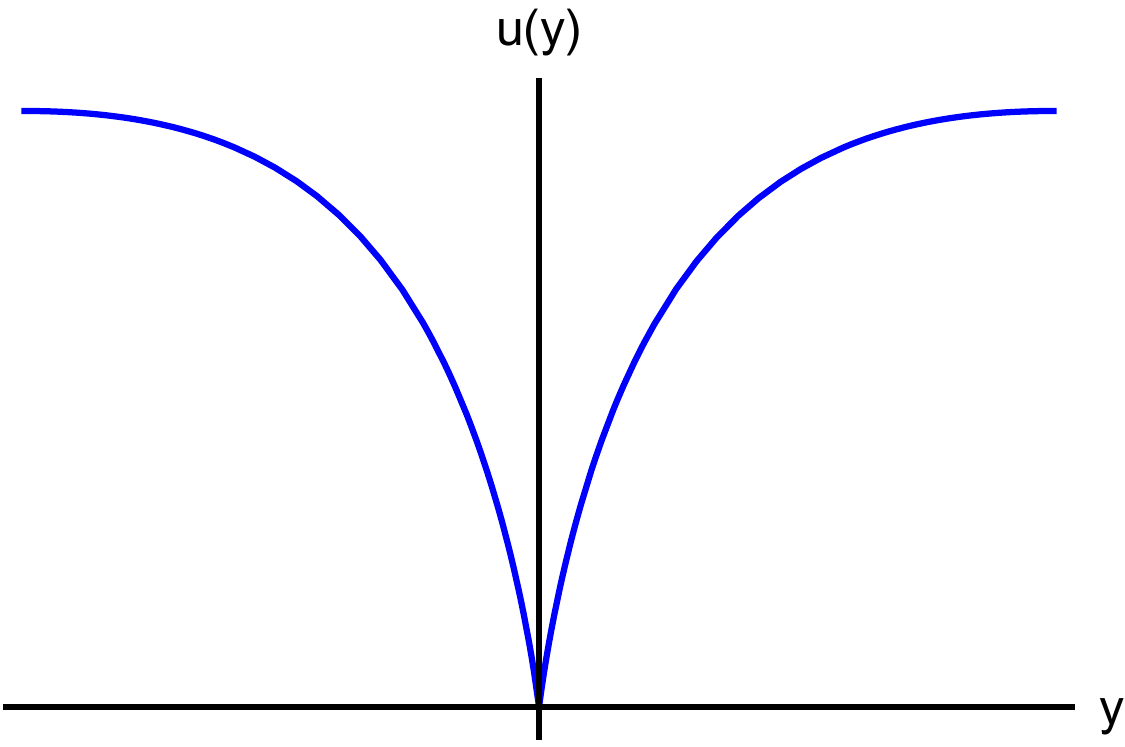} \hspace{1cm}
\includegraphics[scale=0.45]{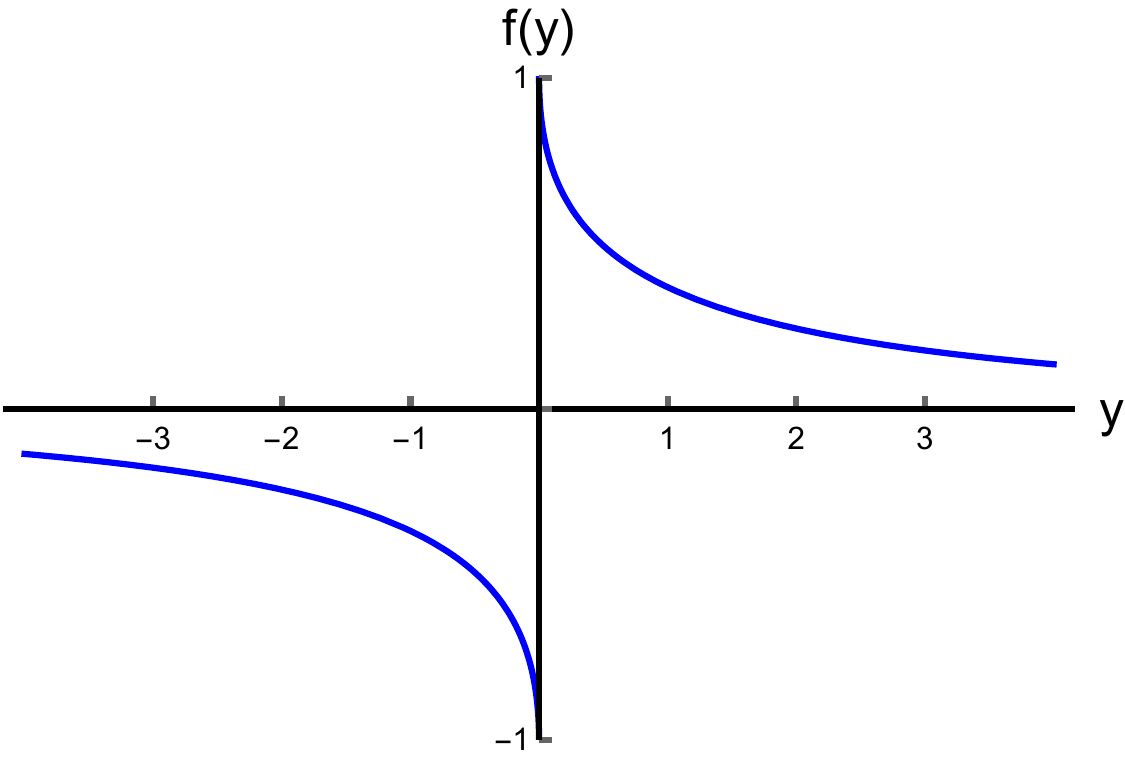}
\end{center}
\caption{Qualitative behaviour of the functions $u(y)$ (left) and $f(y)$ (right)}
\label{sol}
\end{figure}

To be more specific, we considered two cases: $p=5$ and $p=6$.  We solved the corresponding equations for $u(y)$ (\ref{uequation}) numerically in two different ways: we used the Euler method for a qualitative study of the behaviour of the function, the Runge-Kutta 4th order method and the automatic Wolfram Mathematica methods~\cite{Mathematica} for checking the results.  Then we built the function $f(y)$ and substituted it into eq (\ref{sigma}). What is left is to replace the argument 
$$y\to -\frac{g}{16\pi^2}\phi^{p-4}\log{\frac{g\phi^{p-2}}{\mu^2/(p-2)!}}.$$
Now one can plot the effective potential. The results for $p=5$ and $p=6$ are shown in Figs.\ref{3EffPot5} and 
\ref{3EffPot6}. 
\begin{figure}[ht]
\begin{center}
\includegraphics[scale=0.45]{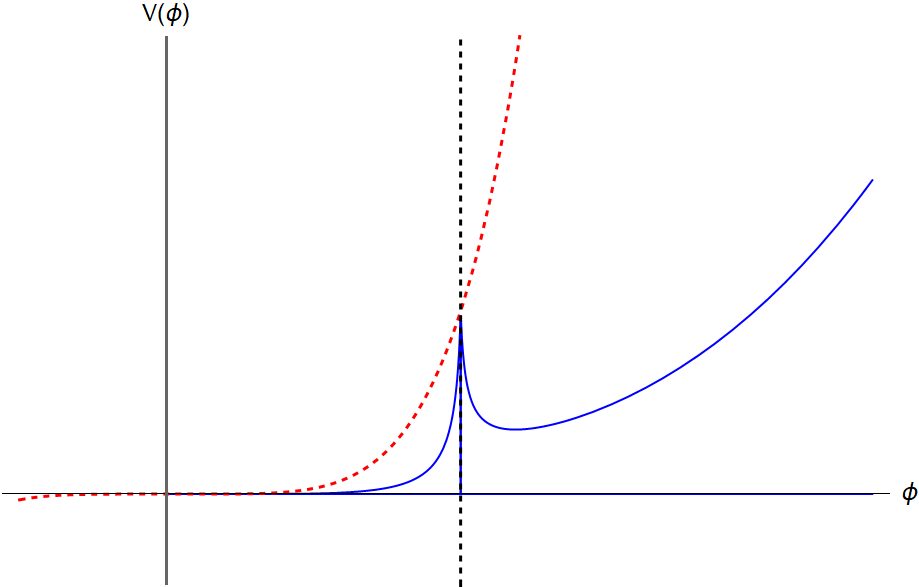} 
\end{center}
\caption{LL effective potential  for the $\phi^5$ theory. The blue line is the LL numerical solution, the red dashed line is the classical potential}
\label{3EffPot5}
\end{figure}
\begin{figure}[h]
\begin{center}
\includegraphics[scale=0.40]{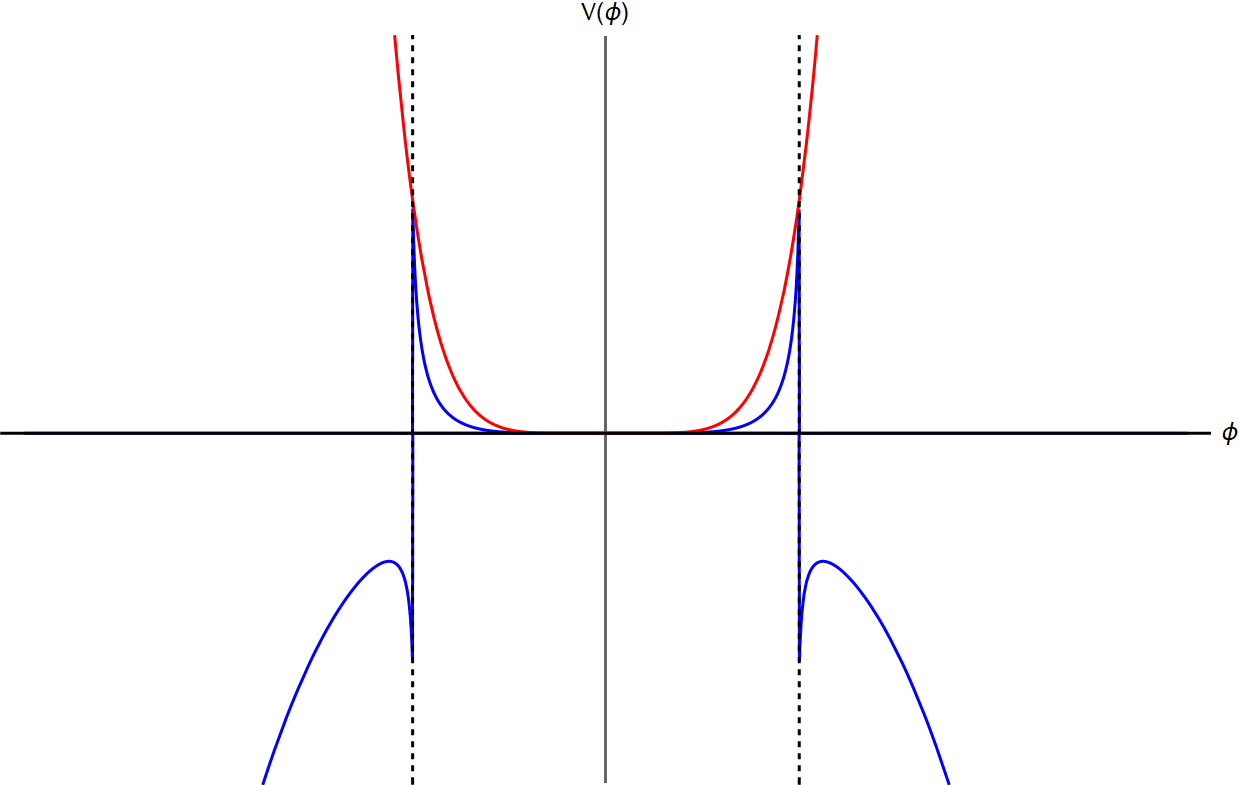} 
\end{center}
\caption{LL effective potential  for the $\phi^6$ (left) theory. The blue line is the LL numerical solution, the red dashed line is the classical potential}
\label{3EffPot6}
\end{figure}

One can see that the behaviour of the effective potential is characterised by the presence of finite discontinuity contrary to infinite or pole in the case of the $\phi^4$ theory. 
The main difference between this discontinuity and the pole one is that the "height" of the potential barrier is finite. So the state located inside the potential is metastable.
In the case of the $\phi^5$ theory, the quantum potential in the LL-approximation has a minimum in the vicinity of potential discontinuity. For a negative field value the quantum potential falls along with the classical potential. 
In the case of the $\phi^6$ theory, the quantum potential in the LL-approximation has no additional minima just like in the $\phi^4$ theory. It seems that this is also true for any even value of $p$.

Once again, one has to check the validity of the approximation. In the case of $p>4$, it is more subtle. For perturbation theory to be valid, one should have 
\beq \frac{g}{16\pi^2}\phi^{p-4} <1, \label{PT}
\eeq
while for the LL approximation to be relevant, one needs 
\beq \log\frac{g\phi^{p-2}}{(p-2)!\mu^2}>1. 
\eeq
This can be achieved for small coupling $g$ and temporary field $\phi$. Take, for instance, the $p=6$ case. 
One has 
\beq
\phi^2<\frac{16\pi^2}{g} \Longrightarrow \log \frac{g\phi^{4}}{4!\mu^2}< \log \frac{(16\pi^2)^2}{24g\mu^2} 
\eeq
The last log term has to be greater than 1, which can be reached for $g\mu^2$ small enough.

At the same time,
the singularity in the effective potential at $y=0$ corresponds to $\frac{g\phi^{p-2}}{(p-2)!\mu^2}=1$. One can see, that this value is compatible  with the requirement (\ref{PT})  if the coupling is small enough. Then the observed discontinuity lies within the validity of PT range.

\section{Example II: Exponential potential}

Another example when the main equation can be reduced to the ordinary differential equation is the exponential potential
\beq
gV_0=ge^{|\phi/m|}.
\eeq
In this case the dimension of the coupling is 4 and the dimensionless variable is $y=z/m^4 e^{|\phi/m|}$. 
Choosing the function $\Sigma(z,\phi)$ in the form
\beq \Sigma(z,\phi)=e^{|\phi/m|} f(y),
\eeq
we obtain the following equation for the function $f(y)$
 \begin{equation}
    f'(y)=-\frac{1}{4} \left(y^2 f''(y)+  3 y f'(y)+f(y) \right)^2
\end{equation}
with the initial condition $f(0)=0, f'(0)=-1/4$.
This equation looks like the one for the power-like potential with $p>4$ (\ref{phiPeq}) and has a similar solution with discontinuity.  The only difference is the value of the effective mass  $v_2 = \frac{g}{m^2} e^{|\phi/m|}$ that leads to the overall substitution  
\beq y \to -\frac{g}{16\pi^2m^4}e^{|\phi/m|}\log  \frac{g}{m^2 \mu^2} e^{|\phi/m|}. 
\eeq
The plot of the effective potential in the LL approximation is shown in Fig.\ref{3EffPotexp}.
\begin{figure}[h!]
 \begin{center}
  \epsfxsize=10cm
 \epsffile{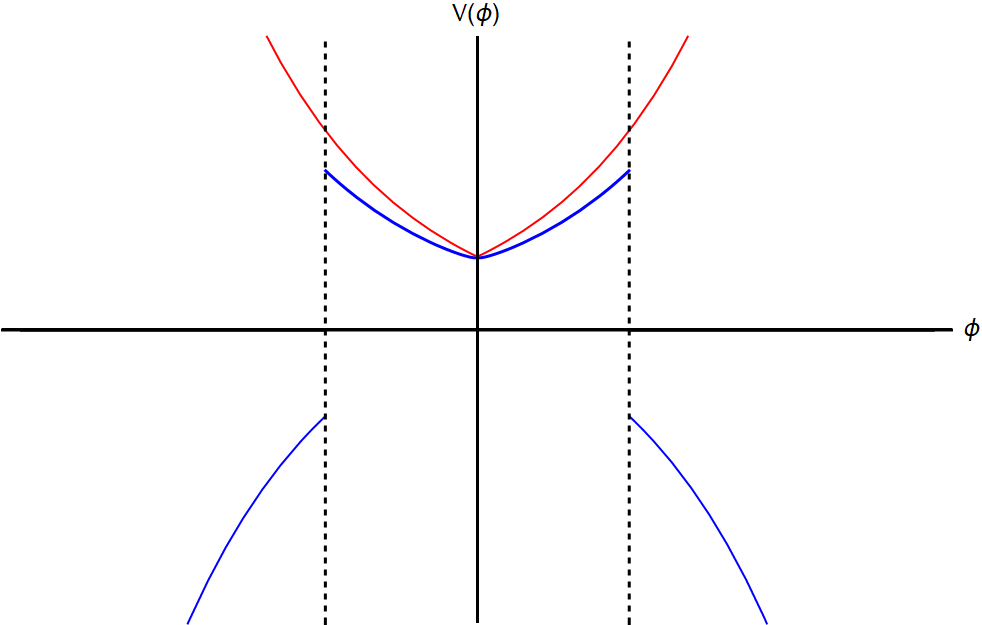}
 \end{center}
 \vspace{-0.2cm}
 \caption{ Effective potential  for the exponential case. The blue line is the LL numerical solution, the red dashed line is the classical potential}
\label{3EffPotexp}
 \end{figure}
 
\noindent Validity of PT now reads
 \beq
 \frac{g}{16\pi^2 m^4}e^{|\phi/m|}<1,
 \eeq
while the application of the LL approximation requires
\beq
\log [\frac{g}{m^2\mu^2}e^{|\phi/m|}]>1.
\eeq
These conditions can be simultaneously satisfied if 
\beq
\phi/m < \log \frac{16\pi^2 m^4}{g} \  \mbox{and} \  \log\frac{16\pi^2 m^2}{\mu^2}>1.
\eeq
The discontinuity point lies within the validity range.

\section{Conclusion}

We conclude that quantum corrections for the effective potential can be calculated for an arbitrary initial classical potential without renormalizability restriction. Of course, this can be done assuming that the UV divergences are removed by some subtraction prescription.  We do not discuss this severe problem here but rather we  derive the consequences of the independence of the leading log approximation of this unknown prescription. Then the effective potential in the LL approximation becomes fully defined and obeys the RG master equation which is a partial non-linear second order differential equation. In some cases, this equation can be reduced  to the ordinary differential one and can be solved at least numerically. We have not considered here other potentials though partial differential equations are eligible to numerical solutions as well.

In all the cases that we have studied above, the obtained ordinary differential equations obey the solution with discontinuity. It seems that this is the general feature of equations like (\ref{phiPeq}). For the effective potential this means that one has a metastable minimum at the origin and no other minima exist. This is in fact similar to the conclusion of the original CW paper that for the pure $\phi^4$ theory the proposed mechanism of radiative symmetry breaking does not work and one has to consider more complicated theories.  

The main message of this paper  is that, under certain assumptions, while studying the CW mechanism, one may not be restricted by renormalizable potentials but consider much wider possibilities. We provided the method of such analysis. This might be useful for cosmological applications where they are usually not limited by renormalizability  since gravity makes it non-renormalizable anyway.

In this paper we considered the effective potential in the case of one field and one coupling. The advocated formalism can be extended to the more complicated cases of several fields/couplings including, for instance, gauge theories like scalar electrodynamics, considered in the original CW paper. The application essentially depends on a physical relevance of a given interaction.

\section*{Acknowledgments}
 Financial support from Russian Science Foundation grant \# 21-12-00129  is cordially acknowledged.

\bibliographystyle{unsrt}
\bibliography{refs}

\end{document}